\documentclass[draft]{agujournal2019}
\usepackage{url} 
\usepackage{lineno}
\usepackage[finalnew]{trackchanges} 
\usepackage{soul}
\draftfalse

\journalname{Geophysical Research Letters}

\begin{document}

%
%


\title{Alternating north-south brightness ratio of Ganymede's
  auroral ovals: Hubble Space
  Telescope observations around the Juno PJ34 flyby}

\authors{
Joachim Saur\affil{1}, 
Stefan Duling\affil{1}, 
Alexandre Wennmacher\affil{1}, 
Clarissa Willmes\affil{1},  
Lorenz Roth\affil{2}, 
Darrell F. Strobel\affil{3} ,
Fr\'ed\'eric Allegrini\affil{4,10}, 
Fran Bagenal\affil{5}, 
Scott J. Bolton\affil{4}, 
Bertrand Bonfond\affil{6}, 
George Clark\affil{7}, 
Randy Gladstone\affil{4,10}, 
T.K. Greathouse\affil{4}, 
Denis C. Grodent\affil{6}, 
Candice J. Hansen\affil{8},
W.S. Kurth\affil{11},
 Glenn S. Orton\affil{9},
Kurt D. Retherford\affil{4,10},
Abigail M. Rymer\affil{7},
A.H. Sulaiman\affil{11}
}

\affiliation{1}{University of Cologne, Institute of Geophysics and Meteorology,  Cologne, Germany}
\affiliation{2}{KTH, Royal Institute of Technology,  School of Electrical Engineering, Stockholm, Sweden}
\affiliation{3}{Johns Hopkins University, Baltimore, MD, USA} 
\affiliation{4}{Southwest Research Institute, San Antonio, TX, USA}
\affiliation{5}{University of Colorado, Boulder, CO, USA}
\affiliation{6}{Universit\'{e} de Li\`{e}ge, LPAP - STAR Institute,
  Li\`{e}ge, Belgium}
\affiliation{7}{Applied Physics Laboratory Johns Hopkins University, Laurel, MD, USA} 
\affiliation{8}{ Planetary Science Institute, Tucson, AZ, USA}
\affiliation{9}{Jet Propulsion Laboratory, California Institute of Technology, Pasadena, CA, USA}
\affiliation{10}{Department of Physics and Astronomy, University of Texas at San Antonio, San Antonio, TX, USA}
\affiliation{11}{Department of Physics and Astronomy, University of Iowa, IA, USA}

\correspondingauthor{Joachim Saur}{saur@geo.uni-koeln.de}


\begin{keypoints}
\item HST observations of Ganymede's orbitally trailing hemisphere on
 June 7, 2021 in support of Juno flyby
\item Brightness ratio of northern and
  southern ovals oscillates such that the oval facing the Jovian plasma sheet
  is brighter 
\item Oscillation suggests the aurora is controlled by
 magnetic stresses coupling the moon's magnetic field to the surrounding Jovian plasma sheet
\end{keypoints}

%
%

%
%



\begin{abstract}
 We report results of Hubble Space Telescope observations from
 Ganymede's orbitally trailing side which were taken around the flyby of the Juno spacecraft on June 7, 2021. 
We find that Ganymede's northern and southern auroral ovals alternate in brightness such that the
oval facing Jupiter's magnetospheric plasma sheet is brighter than
the other one. 
This suggests that the generator that powers Ganymede's aurora is the
 momentum of the Jovian plasma sheet north and south of Ganymede's
magnetosphere. Magnetic coupling of Ganymede to the plasma sheet above
and below the moon
causes asymmetric magnetic stresses and electromagnetic
energy fluxes ultimately powering the auroral acceleration process.
No clear statistically significant 
timevariability of the auroral emission on short time scales of 100s
could be resolved. We show that electron energy fluxes of several tens
of mW m$^{-2}$  are required for its OI 1356
\AA$\;$ emission making Ganymede a very poor auroral emitter.
\end{abstract}

\section*{Plain Language Summary}
Jupiter's moon Ganymede is the largest moon in the solar system and
the only known moon with an intrinsic magnetic field and two auroral
ovals around its north and south poles. Earth also possesses two auroral
ovals, which are bands of emission around its poles. This emission is
also referred to as northern and southern lights.  We use the Hubble Space
Telescope to observe Ganymede's aurora around the time when
NASA's Juno spacecraft had a close flyby at Ganymede. We find that the
brightness of the northern and southern ovals alternate in intensity
with a period of 10 hours.  Additionally, we derive that an energy
flux of several tens of milli-Watt per square meter is necessary to
power the auroral emission. This energy flux comes from energetic
electrons accelerated in the vicinity of Ganymede.

 \section{Introduction}

With its intrinsic magnetic field and two auroral ovals, 
Jupiter's moon Ganymede resembles other planets with intrinsic magnetic fields
and auroral emission \cite{kive96,kive98,hall98,feld00a}. The key difference
compared to magnetized planets is that Ganymede's 
mini-magnetosphere is located within Jupiter's gigantic magnetosphere and  does not possess a
bow shock. However, many processes of Ganymede's unique
magnetosphere such as the mechanisms that power its aurora are still
poorly understood.
The return of the  Juno spacecraft to Ganymede on
\change{June, 7 2021,}{June 7, 2021,}
twenty
years after the Galileo mission thus provides a great opportunity to
further our understanding of the moon.  Juno passed Ganymede on its  34\textsuperscript{th} 
perijove orbit (PJ34) across its orbitally leading side, i.e.,
the downstream side of its magnetosphere. \remove{, a geometry not explored by
  Galileo.}
In support of the Juno
flyby, HST observations with the Space Telescope Imaging Spectrograph (STIS) camera were obtained. HST
observed Ganymede
\change{with}
{on}
three HST orbits before and three orbits after Juno's closest
approach to monitor the evolution of the aurora.  Simultaneous observations
\change{where}{were}
not possible due to scheduling constraints of HST as detailed in
Section 2.

Ganymede's auroral emission was detected with
HST/GHRS \cite{hall98} and provided the first
evidence that Ganymede possesses a thin molecular oxygen atmosphere. A
recent analysis of multiple sets of HST data showed that Ganymede
additionally harbors a sublimation-driven localized water atmosphere \cite{roth21}. 
Spatially resolved observations with HST/STIS have demonstrated that
Ganymede's emission primarily stems from two auroral
structures  around its north and south
poles \cite{feld00a,mcgr13}. The overall brightness and the
location of the aurora changes periodically with Ganymede's position in Jupiter's magnetosphere
\cite{saur15,musa17}.  The total brightness is maximum when Ganymede is
in the plasma sheet of Jupiter's magnetosphere. 
Based on \citeA{mcgr13}, and \citeA{grea22}, 
the auroral ovals match to
regions near the open-closed field line region of Ganymede's
mini-magnetosphere and  delineate locations where the transition from
closed to open field lines of Ganymede's mini-magnetosphere occurs.
Along these field line regions a yet unidentified mechanism accelerates
particles that precipitate into Ganymede's tenuous atmosphere and excite the observed
emission. Various possible mechanisms are discussed in the literature \cite<e.g.>[]{evia01}.
For example, acceleration might be due to reconnection, i.e., merging of Jupiter's and
Ganymede's magnetic field lines at the open-closed field line boundary. This process converts
magnetic field energy into kinetic energy of the
particles. Alternatively, field-aligned electric currents
connecting  into Ganymede's ionosphere might be related to particle acceleration.
Another possibility
would be that electromagnetic waves accelerate
particles which appears
to be the dominant acceleration mechanism for  Jupiter's main aurora
\cite{mauk17,saur18a}. 
A basic understanding of the aurora is of fundamental interest, but it is also important because
the dynamics of its auroral ovals have been used to deduce a subsurface ocean within
Ganymede \cite{saur15}. A better understanding of Ganymede's auroral
emission will also provide important information for the science planning of ESA’s JUICE mission, which will orbit
Ganymede starting 2032 \cite<e.g.,>{gras13}.

\section{Data and data processing}
The HST observations were acquired in support of the Juno flyby and
monitored Ganymede's orbitally trailing side 
on June 7, 2021. The observations were carried out with HST/STIS with grating G140L and aperture
52X2. Details of the exposures are given in Table S1 \add{in Supporting
Information.} 
STIS provides spectral images of Ganymede's emission at several
different wavelengths. In this work we analyze Ganymede's auroral OI
emission at 1356 \AA$\;$
because Ganymede's auroral emission at this wavelength is about a
factor of two brighter than the other oxygen FUV OI emission at 1304
\AA$\;$ \cite{feld00a}. 
At OI 1304 \AA$\;$ in addition to auroral emission there is
solar-resonance 
scattering and surface-reflected emission as well. OI 1356
\AA$\;$  is a semi-forbidden multiplet and hence an optically thin
emission, almost entirely auroral, leading to a superior signal to
noise ratio (SNR) from OI 1356 \AA$\;$  in  comparison to OI 1304 \AA. 
The HST data were processed identically to previous analyses of Ganymede's
observations obtained with STIS \cite<e.g.,>{saur15,musa17}.  The main steps are:
Determination of the location of Ganymede within each spectral image,
removal of the solar reflected light from the surface and rotation of
the resultant spectral image such that Jupiter north is up.

The observations consist of two visits with three HST orbits each (see
Table S2). 
  \begin{figure}
    \hspace*{-3cm}\includegraphics[width=14cm,angle=-90]{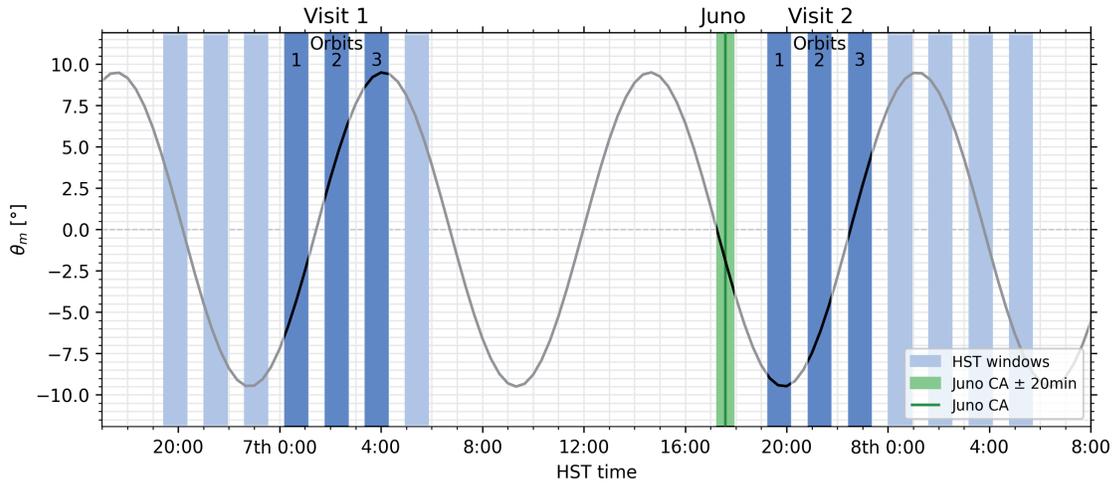}
 
\vspace*{-3.8cm}\caption{Schedule of HST observations with respect to Juno's Ganymede
  flyby. Juno's
  closest approach to Ganymede occurred on June 7, 2021 at 17:35
  (UTC). Light
  blue areas display time windows when Ganymede is observable by HST. 
Executed HST observations (dark blue) consist of three orbits labeled 1, 2,
  and 3 in visit 1
  before the flyby and three orbits labeled 1, 2, and 3 in visit 2 after the
  flyby. 
Magnetic latitude $\Psi_m$ of Ganymede in Jupiter's magnetosphere is
shown as gray curve.}
\label{f_setup}
\end{figure}
In Figure \ref{f_setup}, we display the scheduling of the HST
observation with respect to the Juno flyby at Ganymede. Juno's closest
approach occurred at 16:56 UTC on  June 7, 2021  in
spacecraft event time which corresponds to 17:35 UTC Earth-received
time
\change{considering,}{considering}
the light
travel time of 39 minutes. HST can only observe Ganymede when the
moon is not occulted by Earth and when the South Atlantic Anomaly does
not  impede HST observations, which unfortunately
occurred during an $\sim$14 hr time window around the Juno flyby.  
The resulting available HST windows are shown in
light blue in Figure \ref{f_setup}. 
Ganymede's latitudinal
position $\Psi_m$ with respect to the center of Jupiter's
magnetospheric plasma sheet is
displayed as grey curve
\add{
  The magnetic latitude} $\Psi_m$
\add{is calculated based on a Jupiter
centered tilted dipole model with}
$\Psi_m=9.5^\circ \cos (\lambda_{III}-200.8^\circ)$, where
$\lambda_{III}$
\add{
describes the system III longitude of Ganymede} \cite{dess83,conn98}.
Out of the available windows we scheduled the first three HST
orbits directly after the Juno flyby. They are labeled
orbit 1, 2, and 3 of visit 2 and started 2 hr 9 min after closest approach. Orbit 1 of visit 2 occurs when
Ganymede is near maximum southern latitudes and then Ganymede moves towards the
center of the plasma sheet and crosses it during orbit 3.  Visit 1 was
chosen to be symmetric to visit 2 in the sense that Ganymede covers
approximately the same
\change {distance range}{angular span}
$|\Psi_m|$
\add{with respect}
to the plasma sheet.
The reason is to better compare the
emission pattern before and after the flyby. Observations during visit 1 ended
13 hrs and 13 min before the Juno flyby.

\section{Results}
\begin{figure}
    \includegraphics[width=16cm]{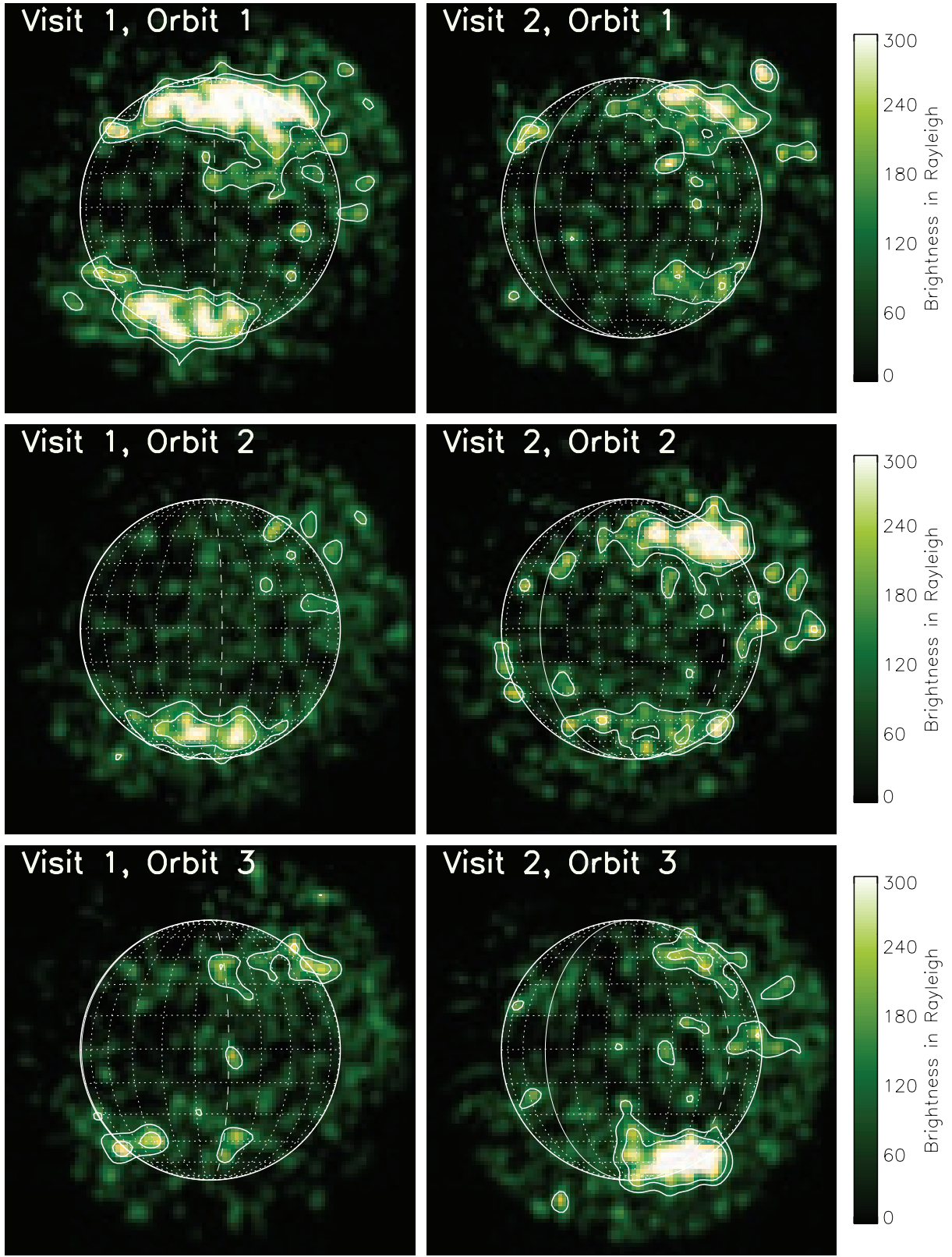}
 \caption{HST/STIS images of Ganymede's auroral brightness in Rayleigh
   at OI 1356
   \AA. Visit 1 occurred before and visit 2 after the
   Juno flyby on June, 7 2021. Observations show mostly Ganymede's
   trailing, i.e., plasma flow upstream, hemisphere. The dashed line indicates the 90$^\circ$ meridian.}
\label{f_obs}
 \end{figure}

Ganymede's auroral emission at OI 1356 \AA$\;$ for the three orbits of
both visits is displayed in Figure
\ref{f_obs}. It demonstrates that the auroral emission on Ganymede's
trailing hemisphere is primarily
located at large northern and southern latitudes as known from
previous observations \cite{feld00a,mcgr13,musa17}. 

\subsection{Context for Juno}
The HST observations taken before and after the Juno flyby provide
context for the observations obtained by
Juno. For each HST orbit, the disk averaged brightness of the OI 1356 \AA$\;$ emission, 
including limb emission up to 200 km above the disk, is summarized in Table S1.
The emission is brightest when Ganymede 
is near the center of Jupiter's magnetospheric
plasma sheet, which occurs
during orbit 1 of visit 1 (about 16 hr before the Juno flyby)
and during orbit 3 of visit 2 (about 5 hr after the flyby).  Ganymede
was also near the center of plasma sheet during the Juno flyby. 
Orbit 1 of visit 1 displays an exceptional high
brightness of 128.8 $\pm$4.1 R, while during orbit 3 of visit 2 
the brightness of 76.2 $\pm$2.6 R was typical, as compared to previous
observations under similar conditions \cite{musa17}.
Ganymede was not observed with HST during the
Juno flyby, but the HST observations with the smaller
temporal separation of 5 hr compared to 16 hr might more likely
represent the state of its auroral emission while the flyby occurred.

\subsection{Alternating north-south brightness ratio of auroral ovals}
During orbit 1 of visit 1, Ganymede was below (i.e., south of) the
plasma sheet (Figure \ref{f_setup}). The HST
observations displayed in Figure \ref{f_obs}, left, reveal that the northern
oval, which faces the center of the plasma sheet, was brighter than the
southern oval. During orbit 2 of visit 1 Ganymede had moved above the plasma sheet and the southern
oval, which then faced the center of the plasma sheet turned brighter. In the last orbit
of visit 1, the emission is patchy and no hemisphere is clearly
brighter than the other one.  During orbit 1 and 2 of visit 2
Ganymede was below the plasma sheet and the
northern oval was brighter (Figure \ref{f_obs}, right). During orbit 3 of
visit 2 Ganymede had moved above the plasma sheet and the southern oval turned
brighter. This pattern demonstrates that the oval which faces the center of
the plasma sheet is brighter than the oval which faces away from the plasma sheet.

\begin{figure}
   \includegraphics[width=10cm]{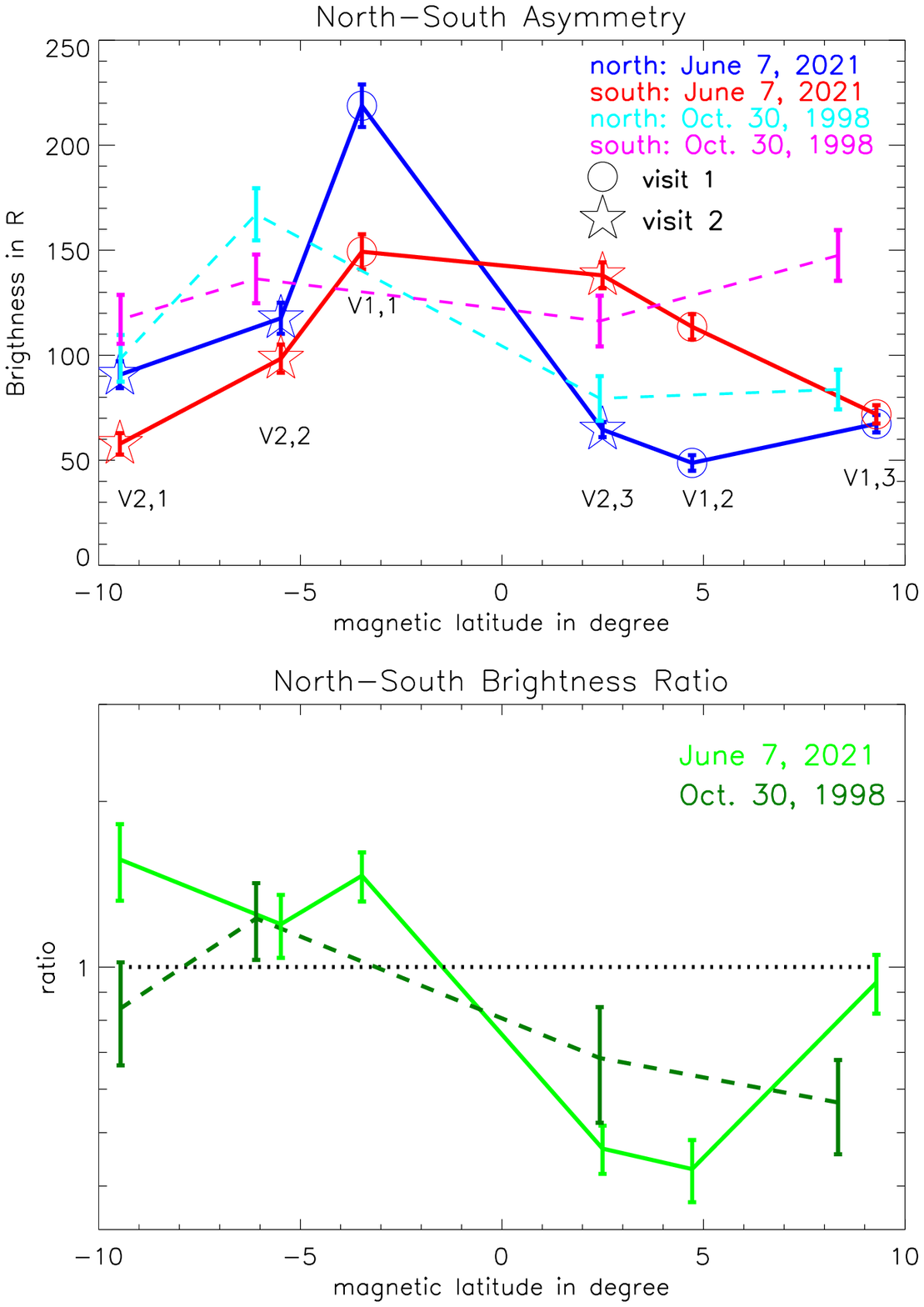}
\caption{North-south asymmetry of Ganymede's auroral emission. Top
  panel: Average brightness of northern and southern
  hemispheres calculated above and below 30$^{\circ}$
latitude within a disk which includes the limb emission within 200
km. The abbreviations V1,1, V1,2, ... V2,3 refer to visit 1 orbit 1, visit 1 orbit 2, ....
  visit 2 orbit 3. Bottom panel: North-South brightness ratio.}
\label{f_asym}
\end{figure}

The brightness of the northern and the southern ovals can be
quantified by the average emission above $30^{\circ}$ and below $-30^{\circ}$
latitude calculated within a disk which includes limb emission within 200
km above the surface.  The resultant average hemispheric brightness is displayed in
Figure \ref{f_asym}
\add{(for details on error bars see Supporting Information).}
The northern and southern brightness as a function of
Ganymede's position with respect to the plasma sheet (approximated by Jovian magnetic
latitude $\Psi_m$) is shown in red and blue, respectively. The northern oval is brighter than
the southern one when Ganymede is below the plasma sheet and vice
versa when Ganymede is above. 
In the lower panel of Figure \ref{f_asym} we display the
brightness ratio between north and south. The ratio with their error
bars demonstrates that the asymmetry pattern is significant. 

Ganymede's trailing side had been observed previously by HST/STIS on October
30, 1998 \cite{feld00a}.  The north-south asymmetry of
these observations is shown in Figure \ref{f_asym} as turquoise and
pink lines in the top panel and the associated ratios in dark green in
the lower panel. Out of the four HST orbits in the 1998 data set, three
orbits follow the same asymmetry trend as the 2021 data while one
orbit at large negative latitudes shows no asymmetry within the error
bars (see left dark green
data points in Figure \ref{f_asym}). These values are consistent with the
analysis of quadrants as function of $\Psi_m$ in \citeA{musa17}. Generally speaking, the 1998 data 
supports the conclusion that  the north-south brightness ratio alternates with Ganymede's
position in Jupiter's plasma sheet such that the hemisphere facing the center
of the plasma sheets is brighter than the opposite hemisphere. The
associated period is the synodic rotation period of Jupiter in the
rest frame of Ganymede, which is
10.54 hr.

\subsection{Short time scale variability}

The STIS  images of Ganymede's auroral emission in Figure \ref{f_obs}
are patchy and thus appear to represent small-scale temporal
variability.  
Therefore we investigate if temporal variability on time-scales
shorter than an HST orbit of
$\sim$40 minutes can be resolved. 
  \begin{figure}
    \includegraphics[width=10cm]{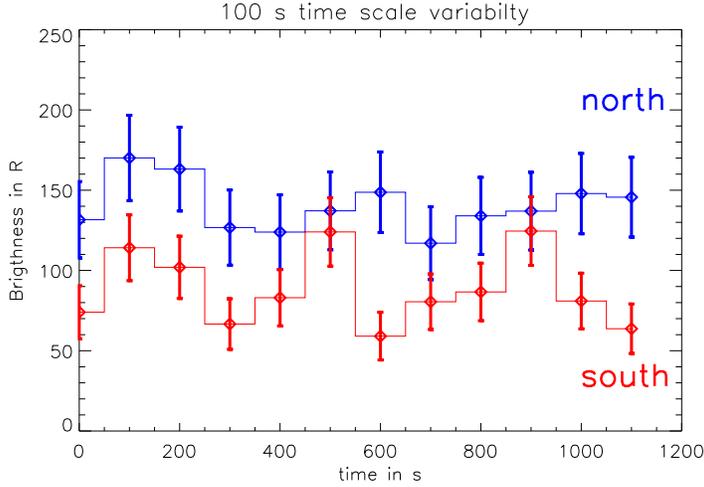}
\vspace*{-3cm}
\caption{Short time scale analysis of intensities measured during orbit 1 of visit 1 on June 7,
  2021. The northern and southern ovals are analyzed separately.}
\label{f_time}
\end{figure}
In Figure \ref{f_time}, we show the brightness within
sub-exposures of 100 s length. 
The brightness is calculated above
$30^{\circ}$  and below
$-30^{\circ}$ latitudes within a disk including 200 km above the limb.
The sub-exposures displayed in Figure \ref{f_time} are from the HST orbit with the brightest
emission of the June 7, 2021 data, i.e.,  orbit 1 of visit 1 (top
left in Figure \ref{f_obs}). Some of the brightness values $b_i$ of
sub-exposures $i$
differ beyond the one-sigma error bars. This is, however, not
sufficient 
to demonstrate true time-variability. For a simple
test we calculate the standard deviation $\sigma_{\mathrm{sub-exp}} = [1/(N-1) \sum
(\bar{b} - b_i)^2]^{1/2} $ of
the sub-exposures (with $\bar{b}= 1/N \sum_N b_i$) and
compare it with the averaged one-sigma uncertainty $\bar{\sigma}_{\mathrm{indiv}} = 1/N \sum_N \sigma_i$
of each of the N sub-exposures $i$. Real time-variability would be
indicated if $r = \sigma_{\mathrm{sub-exp}} / \bar{\sigma}_{\mathrm{indiv}} > 1$,
i.e., the variability between the sub-exposures is larger than their
individual error bars. 
The brightness evolution of the northern ovals in Figure \ref{f_time} is characterized by
$r = 0.84$ indicating that the observations are consistent with a
time-constant emission. The southern oval has a ratio of $r = 1.1 $
indicating possible time-variability. However, 
averaging over all six HST orbits we find $r = 0.91$ for the northern oval and 0.99 for the
southern oval. Detailed statistical values for all orbits can be
found in Table S2 in Supporting Information. The $r$-values thus imply 
no clear sign of short-term variability of 100 s. Further, more
sophisticated tests are required to derive stronger conclusions.

\subsection{Auroral brightness and required electron energy fluxes}
An auroral brightness on Ganymede of 100 R at OI 1356 \AA$\;$ can be explained following 
\citeA{evia01}  by an electron population with surface density of 100
$\times 10^6$ 
m$^{-3}$ and temperature of 100 eV within an O$_2$ column density of
$3 \times 10^{18}$ m$^{-2}$
\add{(see Figure S1 in Supporting Information)}.
\add{For intense brightness values of 1000 R such as seen
  by Juno} \cite{grea22},
\add{and electron density with 950
$\times 10^6$  m$^{-3}$ and a temperature of 200 eV are required (Figure S2).
}  
Assuming a Maxwellian distribution, we can
calculate the associated energy flux in
one-direction as $ (4 \pi)^{-1/2} \;  m_e n_e (2 k_B
T_e /m_e)^{3/2}$, which assumes 5.4 mW m$^{-2}$
\add{
for the 100
R and 
140 mW m$^{-2}$ for the 1000 R cases, respectively}
(with the
electron mass $m_e$ and the Boltzmann constant $k_B$). 
\add{Up to a neutral column density of $3 \times 10^{19}$ m$^{-2}$ the
atmosphere is 'optically thin' for electrons and thus a 10 times smaller electron
density within a ten times larger neutral column density leads to the
same brightness values.  }
Thermal ionospheric electrons are expected to be much colder and
electrons and the neutrals are approximately collisionless in
Ganymede's atmosphere and ionosphere \cite{evia01}.  Electrons within
the downward going loss cone will thus be lost to the surface as seen in
JADE and JEDI data \cite{alle22,clar22}. 
Therefore an energy flux from the magnetosphere into the
atmosphere/ionosphere 
on the order of \add{10 to 100} mW m$^{-2}$ is required to produce the brightnesses
of Ganymede's aurora. The electrons need to be accelerated in the
vicinity of Ganymede since the electron energy density in Jupiter's
magnetosphere near Ganymede is insufficient to produce the aurora \cite{evia01}.

The brightness of Jupiter's main aurora is on the order of several 100
kR. The energy fluxes required to produce 100 kR at Jupiter are $\sim$10 mW m$^{-2}$
\cite{mauk17,gust16}. 
Thus for similar energy fluxes, the
efficiency to produce auroral emission in Jupiter's atmosphere is 
\add{on the order of 10$^3$ times larger} than in
Ganymede's atmosphere. This difference can alternatively be
demonstrated by considering Ganymede's FUV luminosity  of $\sim$3 $\times$ 10$^7$ W
\cite{saur21}. Assuming the emission originates from 10\% of
Ganymede's area, the emitted energy flux is 3.5 $\times$
10$^{-6}$ W m$^{-2}$. For electron fluxes on the order of 10 to 100 mW
m$^{-2}$, this would correspond to an efficiency to convert electron
\add{
  energy into FUV emission of
}
$\sim$10$^{-4}$,
\add{
  which is a factor of
}
10$^2$ to 10$^{3}$
\add{
  smaller
compared to the FUV efficiency of Jupiter's main
aurora  in the range
}
10$^{-1}$   to 10$^{-2}$ \cite{bhar00,mauk17}.
The primary reasons for the difference are: 
(1) Jupiter's atmosphere is collisionally thick and energetic
electrons deposit all of their energy in the upper atmosphere
while Ganymede's atmosphere is collisionally thin and most electrons
are lost to the surface of Ganymede. (2) In Jupiter's hydrogen atmosphere
about 10 - 20\% of electron energy is radiated in FUV H$_2$ band
emissions, but only about 1\% of the energy of electrons interacting
with Ganymede's oxygen atmosphere is radiated away as OI 1356
\AA$\;$  photons and most of the energy is spent in ionization and
dissociation of O$_2$.

\section{Conclusions and discussion}
HST observations of Ganymede taken on June 7, 2021 provide context for the 
flyby of the Juno spacecraft at Ganymede.  
Based on the HST observations closest to the flyby, Ganymede's auroral emission and thus its
magnetospheric particle environment appear to have  been in
a typical state (i.e., similar to previous
observations) during the Juno flyby.

\change{We find}{The analysis of the HST observations for Ganymede's
  trailing hemisphere shows} that the brightness ratio of the northern and southern ovals
oscillate such that the hemisphere facing Jupiter's plasma sheet is
brighter. This suggests that the properties of the surrounding plasma sheet
in the northern and southern direction control the brightness of
Ganymede's aurora. In case of Io and Europa, an asymmetry in the
north-south limb UV emission was observed as well
\cite{reth03,roth14a,roth16}. The underlying reason for the asymmetry
in case of Io and Europa was attributed to the larger electron thermal
energy reservoir on the side facing the surrounding plasma sheet. However, in case of
Ganymede the cause of the asymmetric auroral emission cannot be
attributed to the properties of Jupiter's magnetospheric
electrons. Due to the low plasma
density at Ganymede's orbit, these magnetospheric electrons do not possess a
sufficiently large energy density to cause Ganymede's aurora
\cite{evia01}. Therefore other energy fluxes into Ganymede's magnetosphere
and a local electron energization process 
are needed in contrast to Europa
and Io where the electron energy density in the surrounding plasma is
sufficient to power their auroral emission. 

The plasma density in Jupiter's plasma sheet is the property which primarily
changes in the north-south direction above and below Ganymede, while the plasma velocity is
relatively similar \cite{bage11}. The gradient of the plasma density
results in a larger plasma momentum on the side facing the
center of the plasma sheet. Close to Ganymede, the plasma is slowed due to the
interaction of Jupiter's magnetospheric plasma with Ganymede's
internal magnetic field and the collisions with its atmosphere. 
Since Ganymede and the surrounding plasma are connected
by magnetic field lines, the slowed plasma in
Ganymede's vicinity and the fast flowing plasma above and below
Ganymede is generating magnetic stresses around Ganymede. On the side
with the larger momentum, the magnetic stresses are larger. The larger
momentum also causes the Alfv\'en wings
on that side to be more strongly bent back compared to the other
hemisphere. The bent-back angle is proportional to the square root of the plasma density
\cite{neub80,saur13}. 
The larger bend and the larger stresses need to be maintained by electromagnetic
energy fluxes between the plasma sheet and Ganymede. These energy fluxes thus
appear to be the root power source for the auroral emission. 
The asymmetric stresses are also the cause for larger electric currents
feeding into Ganymede's magnetosphere on one hemisphere. However, the electric currents
enter the Ganymede system primarily on the flanks of Ganymede and not
on the upstream and downstream side where auroral emission is observed
\cite{neub98,jia08,duli14,duli22}.  Reconnection might also be
asymmetric within an inhomogeneous background plasma but the scales of the
gradient of the background properties are large compared to the size of
Ganymede. We therefore
suggest that the asymmetric plasma momentum causes asymmetric  magnetic stresses with associated
asymmetric electromagnetic energy fluxes towards Ganymede.
These fluxes ultimately provide the power for Ganymede's auroral acceleration
processes and its auroral emission.

Our simple analysis of 100 s sub-exposures reveal no 
statistically significant  time-variability of the total auroral brightness
on short time scales.  However, further studies with more sophisticated tests and
better resolved observations such as with JUICE \cite<e.g.,>{gras13} are
warranted to detect possible short term variability not resolved in this study.

\section{Open Research}
This work is based on observations with the NASA/ESA
Hubble Space Telescope obtained at the Space Telescope
Science Institute, which is operated by the Association of
Universities for Research in Astronomy (AURA), Inc., under
NASA contract NAS 5-26555.
All data used in this study is available on the Mikulski Archive for
Space Telescopes (MAST) of the Space Telescope Science Institute at
\url{http://archive.stsci.edu/hst/}.  The specific data sets are listed in
Table S1 and can be accessed at:\\
\noindent\url{https:https://mast.stsci.edu/search/ui/#/hst/results?resolve=true&radius=3&radius_units=arcminutes&data_type=all&observations=S&active_instruments=stis,acs,wfc3,cos,fgs&legacy_instruments=foc,fos,ghrs,hsp,nicmos,wfpc,wfpc2&proposal_id=16499&select_cols=ang_sep,sci_aper_1234,sci_central_wavelength,sci_data_set_name,sci_dec,sci_actual_duration,sci_spec_1234,sci_hlsp,sci_instrume,sci_pi_last_name,sci_preview_name,sci_pep_id,sci_ra,sci_refnum,sci_release_date,scp_scan_type,sci_start_time,sci_stop_time,sci_targname&useStore=false&search_key=819f8c5cea95f}.

\acknowledgments
J.S. appreciates the help of William
Januszewski and Joleen Carlberg from STScI in scheduling the observations.
This project has received funding from the European Research Council
(ERC) under the European Union's Horizon 2020 research and innovation
programme (grant agreement No. 884711).
DFS support by STScI HST-GO-16499.006-A grant under NASA contract NAS5-26555.
Some of this research was
carried out at the Jet Propulsion Laboratory, California Institute of
Technology, under contract with NASA (80NM0018D0004).



%
%


%
%
%
%
%

\end{document}


%
%


\title{Supporting Information for "Oscillating north-south brightness ratio of Ganymede's
  auroral ovals: Hubble Space
  Telescope observations around Juno's PJ34 flyby"}
%
%

%
%



\authors{
Joachim Saur\affil{1}, 
Stefan Duling\affil{1}, 
Alexandre Wennmacher\affil{1}, 
Clarissa Willmes\affil{1},  
Lorenz Roth\affil{2}, 
Darrell F. Strobel\affil{3} ,
Fr\'ed\'eric Allegrini\affil{4,10}, 
Fran Bagenal\affil{5}, 
Scott J. Bolton\affil{4}, 
Bertrand Bonfond\affil{6}, 
George Clark\affil{7}, 
Randy Gladstone\affil{4}, 
T.K. Greathouse\affil{4}, 
Denis C. Grodent\affil{6}, 
Candice J. Hansen\affil{8},
W.S. Kurth\affil{11},
 Glenn S. Orton\affil{9},
Kurt D. Retherford\affil{4},
Abigail M. Rymer\affil{7},
A.H. Sulaiman\affil{11}
}

\affiliation{1}{University of Cologne, Institute of Geophysics and Meteorology,  Cologne, Germany}
\affiliation{2}{KTH, Royal Institute of Technology,  School of Electrical Engineering, Stockholm, Sweden}
\affiliation{3}{Johns Hopkins University, Baltimore, MD, USA} 
\affiliation{4}{Southwest Research Institute, San Antonio, TX, USA}
\affiliation{5}{University of Colorado, Boulder, CO, USA}
\affiliation{6}{Universit\'{e} de Li\`{e}ge, LPAP - STAR Institute,
  Li\`{e}ge, Belgium}
\affiliation{7}{Applied Physics Laboratory Johns Hopkins University, Laurel, MD, USA} 
\affiliation{8}{ Planetary Science Institute, Tucson, AZ, USA}
\affiliation{9}{Jet Propulsion Laboratory, California Institute of Technology, Pasadena, CA, USA}
\affiliation{9}{NASA Jet Propulsion Laboratory, Pasadena, CA, USA}
\affiliation{10}{Department of Physics and Astronomy, University of Texas at San Antonio, San Antonio, TX, USA}
\affiliation{11}{Department of Physics and Astronomy, University of Iowa, IA, USA}


%
%

%

\begin{article}

%
%
\end{article}
\noindent\textbf{Contents of this file}
\begin{enumerate}
\item Text S1, S2
 \item Figures S1
\item Tables S1, S2 
\end{enumerate}


\noindent\textbf{Introduction}

The supporting information supplied here provides additional text,
one figure and one table to illustrate in more detail observations and statements made in the main manuscript.

\noindent\textbf{Text S1.}
The essential physics in the calculation of UV emissions from an
atmosphere/exosphere of O and O$_2$ including electron impact, solar
resonance scattering, and solar surface reflection have been discussed
in \citeA{hall95}. The ratio of airglow/auroral emission
rates for OI 1356/OI 1304 exceed 1.3 on Ganymede \cite{hall98}
and are diagnostic that OI 1356 is mostly due to dissociative
electron impact excitation of O$_2$. Our calculations were performed with
the best available measured O$_2$ emission cross sections from
\citeA{kani03}. With these O$_2$ cross sections the emission ratio of
OI 1356 /OI 1304 exceeds 2 over the energy range of these
measurements.  Performing a comparable calculation for the same
emission ratio for electron impact on O atoms yields a ratio that does
not exceed 0.4.  The measured O$_2$ emission cross sections were
extrapolated with the Bethe-Oppenheimer expression $\sigma$1356(E) =
4.05 $\times $10$^{-19}$ E$^{-1}$ ln(44 E) cm$^2$, for E in keV above $>$0.6 keV.  Based on
\citeA{hall98}, we adopt an O$_2$ radial column density of 3 $\times$ 10$^{14}$
cm$^{-2}$, which implies that the exobase is near the surface of
Ganymede.

\begin{figure}
 \noindent\includegraphics[width=16cm]{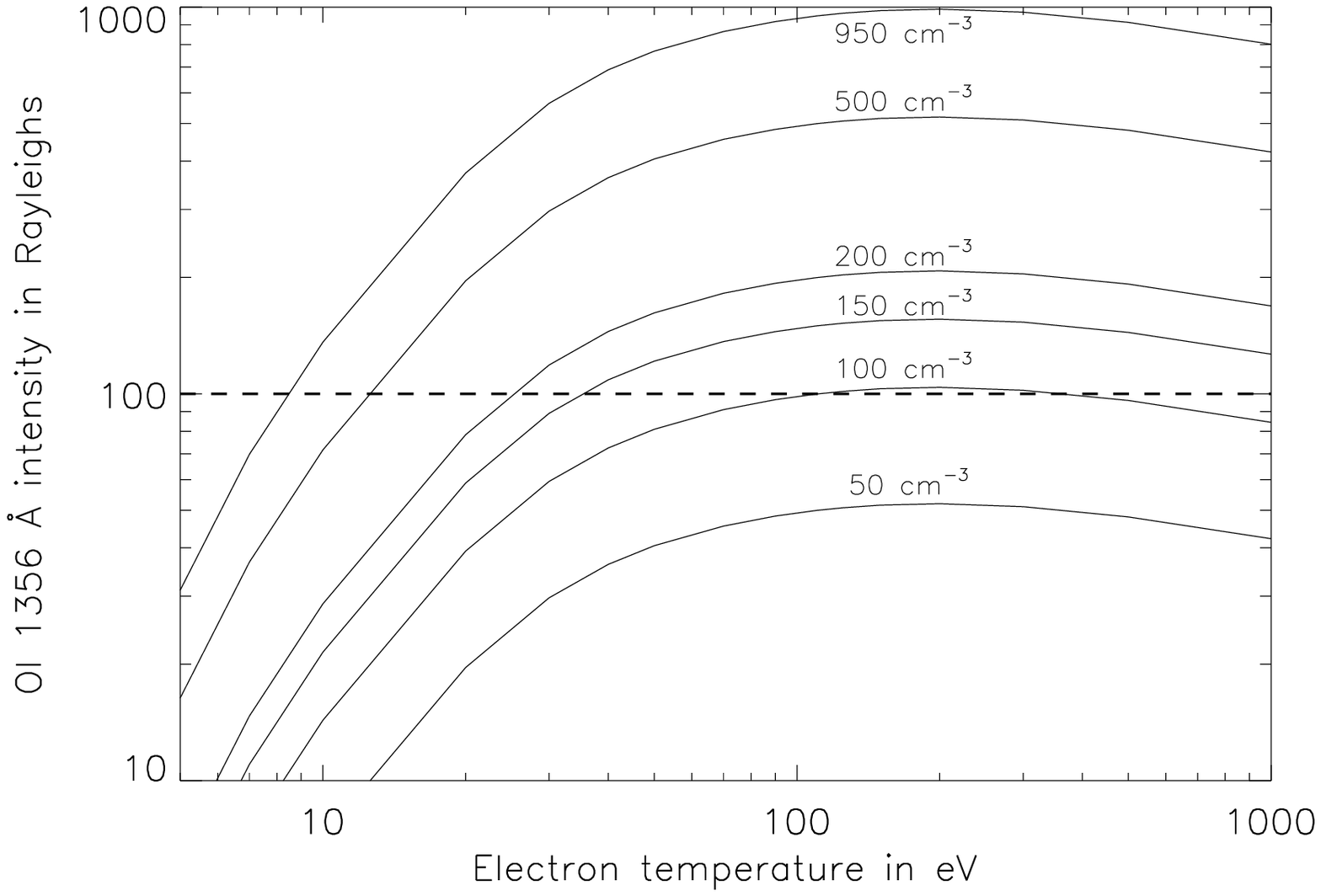}
\caption{Intensity of OI 1356 \AA$\;$ emission as a function of the
  electron temperature in Ganymede's ionosphere/atmosphere for various
  electron densities.
}
\label{f:darrell}
\end{figure}

These calculations with updated O$_2$ cross sections are displayed in
Figure \ref{f:darrell}. They confirm the previous
results in Figure 3 of \citeA{evia01} up to approximately 300 eV
and replace the intensities predicted for electrons accelerated into
the keV energy range due to the extrapolated cross sections based on
the Bethe-Oppenheimer approximation.

\noindent\textbf{Text S2.}
The error bars in Figures 3 and 4 are calculated based on the counts
provided in the flt-files. Each data point in Figure 3 a and Figure 4
is calculated based on the counts  within the respected exposures and the considered
area on Ganymede. The total signal $S = A + R + B$ within the specified
areas of Ganymede is composed of photons from the auroral emission
$A$, photons due to reflected light on the surface of Ganymede $R$ and
due to background noise $B$ of the detector and the sky. The signal to noise SNR
in the Poisson distributed photon fluxes is then given by
$ SNR = (S -R -B)/(S+R+B)^{1/2}$.
The error bars on the ratio of signals in Figure 3 b is calculated
based on standard error propagation \cite{bevi03}.

\newpage
\begin{table*}
      \caption[]{Exposure details of HST/STIS observations of Ganymede in support of the Juno flyby on
        2021-06-07 (ID:
        16499). 
Juno's closest
approach to Ganymede occurred at 17:35 UTC  including the light travel time of
39 min to Earth. Orbit 3 of visit 1 and orbit 1 of visit 2
        were split up into two exposures each for technical reasons.
$^a$: UTC at exposure start (light travel time included), $^b$: magnetic latitude, $^c$: Disk
averaged auroral brightness for OI 1356 \AA$\;$ including limb
emission up to 200 km, normalized
to $\pi (R_G + 200  \mbox{km})^2$, for complete
HST orbits.
}
\label{table:1}      
\centering          
\begin{tabular}{c c c r r r r r  r r r}     
\hline\hline       
Visit & Orbit & Exp & Root name &Obs. Time$^a$ &Exp. Time&$\Psi_m$ $^b$&brightness$^c$
\\ 
\# & \# & \# & & hh-mm-ss & s & degree&R\\
\hline                    
   1 & 1 & 1 & oejj01010  & 00:54:04 & 1631&-3.5 &128.8 $\pm$4.1  \\  
\hline
   1 & 2 & 2 & oejj01020 & 02:29:29& 2291&4.7 & 66.3 $\pm$2.5\\ \hline
   1 & 3 & 3 & oejj01030 & 03:43:16 & 1000 &  9.3& 50.8 $\pm$2.3\\
   1 & 3 & 4 & oejj01040 & 04:04:38 & 1076& 9.5& \\
\hline
   2 & 1 & 5 & oejj02010 & 19:44:20& 866& -9.5&67.3 $\pm$2.9  \\
   2 & 1 & 6 & oejj02020 & 20:01:06& 800 &-9.5&\\               
\hline
   2 & 2 & 7 & oejj2A010 & 21:32:56 & 1514&-5.5&88.2 $\pm$3.5\\
\hline
   2 & 3 & 8 & oejj2A020 & 23:08:23 & 2331&2.5& 76.2 $\pm$2.6\\       
\hline                  
\end{tabular}
\end{table*}

\newpage

\begin{table*}[b]
      \caption[]{ Statistical properties of 100s sub-exposures of all 6
        HST orbits,
        $\bar{b}$ is the average brightness during each orbit,
        $\sigma_{sub-exp} $ is the standard deviation of the brightness
        variability  of
        the sub-exposures during each orbit, and $ \bar{\sigma}_{indiv}$ is
        the average over the individual brightness uncertainties in
        each sub-exposure based on count statistics,  r = ratio of
        brightness variability within sub-exposures to averaged one-sigma
        brightness uncertainty of individual sub-exposures, i.e., for
$ \sigma_{\mathrm{sub-exp}} / \bar{\sigma}_{\mathrm{indiv}} $
                                                                   a
                                                                   value
                                                                   of
                                                                   $r
                                                                    <
                                                                    1$
                                                                    indicates
                                                                    variability
                                                                    could be
                                                                    consistent
                                                                    with
                                                                    statistical
                                                                    noise,
                                                                    only.
}
\label{table:2}      
\centering          
\begin{tabular}{c c c r r r r r  r r r}     
\hline      
\\
\vspace*{0.1cm}
Visit & Orbit & Hemisphere &  $\bar{b}$&  $ \sigma_{\mathrm{sub-exp}} $ &
                                                              $   \bar{\sigma}_{\mathrm{indiv}} $ & 
$r= \sigma_{\mathrm{sub-exp}} / \bar{\sigma}_{\mathrm{indiv}} $\\ 
\# & \# &  &R &R & R & \\ \hline
  1 & 1 & North & 231.1 &39.7  & 47.3 & 0.84 \\ 
  1 & 1 & South & 148.0 &39.1  & 35.5 & 1.10 \\ 
  1 & 2 & North & 42.7 & 32.7 & 30.3 & 1.08 \\ 
  1 & 2 & South & 108.1& 21.5 & 30.7 & 0.70\\ 
  1 & 3 & North & 58.3 &27.6 &  31.7& 0.87 \\ 
  1 & 3 & South & 65.5 &26.5 &  26.3&  1.01 \\ 
  2 & 1 & North & 99.7& 35.1 &  37.4&  0.94\\ 
  2 & 1 & South & 64.0 &28.9 &  23.4&  1.24\\ 
  2 & 2 & North & 118.1&38.1 &  41.1&  0.92\\ 
  2 & 2 & South & 92.4 &28.4 &  29.3&  0.97\\ 
  2 & 3 & North & 54.6 &30.9 &  36.5&  0.85\\ 
  2 & 3 & South & 121.2 &32.9 &  34.6&  0.95\\ 
\hline  
average && North & &34.0 &37.4 & 0.91\\
average && South & &29.5 & 30.& 0.99 \\
\hline               
\end{tabular}
\end{table*}

\newpage